# Analysis of the Institutional Free Market in Accredited Medical Physics Graduate Programs

Brian W. Pogue, Alexander P. Niver

Department of Medical Physics, University of Wisconsin-Madison, Madison WI USA 53705

**ABSTRACT**

**ABSTRACT:**  Medical Physics education is largely delivered through accredited programs where admissions and funding for students is determined by the individual institutions providing the educational experiences. Public data from these accredited graduate programs, along with funding data, can be used to analyze institutionally driven trends in the market for this education.  Temporal trends from 2017 to 2023 show robust growth in MS graduates, increasing at an average of 17.7 per year, as compared to steady but modest growth in PhDs, increasing by 3.6 per year. The current status is that there are nearly two MS graduates for every PhD graduate in North America.  Trends in funding show self-funding of students is a dominant pathway in domestic programs, with this being less dominant in Canadian and Irish programs.   Those programs dominated by accredited MS education have their largest fraction of faculty in radiation oncology departments, whereas those dominated by PhD education have their largest fraction of faculty in radiology departments. Overall NIH funding in the space of radiation diagnostics and therapeutics has been largely static over this timeframe, but with a notable recent rise in NCI funding in the last 5 years.  This can be contrasted to a substantial 5X-6X rise in NIH funding for engineering research programs during this same period, with significant increases in trainee funding there. Taken as a whole, this survey shows that growth in the field of medical physics education is dominated by MS graduates, presumably servicing the expanded growth needs for well-trained clinical physicists.  However, the research infrastructure that supports PhD training in medical physics seems likely to be growing modestly and missing the growth trend of NIH funding that appears to show substantially more growth in non-accredited programs such as biomedical engineering.  This data is useful to informing accreditation guidance on numbers of graduates to match the workforce needs or for inter-institutional planning around education goals.

## 1.0 Introduction

The field of Medical Physics is dominated by the 9000 members of the American Association of Physicists in Medicine (AAPM) [1], and for 3 decades has seen continuous growth at about 200 additional members per year [2-4]. To sustain this growth, there has been annual growth in the number of students trained in medical physics, although matching the needs and supply is a constant challenge. While the structure and content of training is administered by Commission on Accreditation of Medical Physics Education Programs, Inc. (CAMPEP) [5], the growth in supply of trained medical physicists is actually distributed and controlled by each participating educational institution. It is their faculty and governance structures who determine what they want to participate in, and to what extent and who gets admitted and trained. This is the free market of education where faculty and academic administrators make strategic choices about which programs they will offer and support at their own institution and at what scale. The driving factors in education are often quite locally driven, and depend upon their department focus, finances, research capabilities and capacity and sometimes historical experiences. Publicly available data can help assess which factors are most dominant on a national scale, providing a window into the distributed decision making at the institutional level.

The most visible influence on the field of education is the accreditation process from the CAMPEP [7], which formalized requirements for accredited educational programs in 2014, requiring accredited degrees for all applicants to accredited clinical residency programs [8][9]. CAMPEP is aligned with the goals of the American Association of Physicists in Medicine (AAPM) to ensure that the field grows with well-trained individuals [6, 10]. An independent survey of trends is possible because of the CAMPEP requirement that each program post public data on their institutional website, and so this was the source of data for much of this analysis here.

The required didactic training components for accredited programs are prescribed by CAMPEP, with guidelines for admission criteria, curricula, student experience and faculty expertise [6][7], but still admissions and support decisions are completely dependent upon the local institutions. The choices of type of program, MS or PhD, and these opportunities are especially relevant for the balance of education and research but also determine the shape of the departments that support them. Part of the accreditation process requires public posting of data by the universities, which allows for independent analysis of the ongoing trends, as was conducted here. The results shown here might be used to interpret how these free-market forces are influencing the pool of trained medical physicists, who ultimately will be the ones to shape the future of the field.

Inevitably, trends in education are affected by the methods by which student training is funded. A dominant factor at many institutions is the tradeoff in choices of how MS versus PhD students are funded, and so a study of this was conducted across institutions and as an annual aggregate trend. PhD program students are commonly funded, at least in part, by extramural support mechanisms, and so a major part of this is derived by National Institutes of Health (NIH) funding [3]. The NIH funding data is also publicly available on their RePORTER database tool, and so a survey of this was completed to analyze available funds and some measure of successful research by funding training grant awards. The two most prestigious and competitive awards for training PhD students are the F31 Ruth L. Kirschstein Predoctoral Individual National Research Service Award [11], competitively awarded to individual students midway through their training, and the T32 Ruth L. Kirschstein Institutional National Research Service Award, competitively awarded to institutions who create a specific structured training program in a core area for select candidates. Gaining a complete independent picture of funding of programs is a particularly challenging process because this type of data is not easily publicly available from individual programs, although some internal survey data was made available here.

Another piece of data available for this report was the faculty makeup of the program, as broken down by their primary department affiliation. On the surface of it, this is not particularly informative, however aggregate numbers broken into separate groups of institutions was examined to assess trends based upon departmental programs participating and how this may influence choices in MS versus PhD student numbers. Given that clinical versus academic departments may have quite different emphases or funding capabilities, this breakdown provides some insight into graduate program choices.

This aggregate pool of publicly available data provides a snapshot of the field as well as the annual trends in degrees, funding and institutional positioning of education programs, which will all affect the field of Medical Physics. The value of this report is that it provides an independent review of this public data. Most importantly, the numbers of people educated in medical physics is driven by the free markets of individual educational institutions, and so this data can help institutions understand what the landscape is, and where their own emphasis may make the most sense, or how they could adapt to suit their own needs in the future.

**2.0 Methods**

The survey of data started with the required self-reporting totals from each documented CAMPEP graduate program, about their admits and graduates from MS, PhD, Certificate and DMP programs. These data are required to be publicly posted on each accredited program website, and so these data were downloaded for each program. There were 59 CAMPEP accredited graduate programs found online, of which 54 had some information accessible on their websites. Within this, 22 programs did not have complete data posted, and so specific information about certain years student numbers or faculty department information were solicited directly from the program directors by email query. A total of 14 responses were received from these 22, providing us with complete data from 51 programs, and this was the set used here. Of the programs not included, they appeared to have a small number of students, leading to our belief that their omission would not distort the aggregate trends seen here.

The individual graduate program faculty from each program is also available on their websites, and so these were aggregated in terms of which department they belonged to. There were 55 programs included in this faculty analysis, where there was ability to determine the likely home department of each faculty member. Again, for those programs that did not have complete faculty program listing, this information was obtained through email to the program directors. The cumulative data on each home department was thought to be pertinent to understanding the overall structure of the programs that offer the CAMPEP degree.

CAMPEP Program Student support funding info was solicited by CAMPEP and access to this survey data was granted for this research to add in cumulative information about MS and PhD student funding as an aggregate. This information is somewhat limited in value because it does not report quantitative numbers of students supported by each mechanism but rather just listed the range of funding mechanisms possible for each program.

Finally, data on funding for research is available from the NIH in their RePORTER system, which can be queried through the Advanced Search tab, in a number of separate ways with funding year and by institution and department. The data for Medical Physics programs cannot be queried directly but funding for RADIATION-DIAGNOSTIC ONCOLOGY, which is a category that combines departments of Radiology and Radiation Oncology together.  This is an unfortunate grouping that cannot be easily separated out for further insight, but was used here nonetheless because it does cover the majority of departments in which CAMPEP graduate programs find themselves. The other relevant grouping that is easy to query is Physics and Engineering, and so these were surveyed as well. The total number of grants funded each year in each grouping was queried as well as training grants that include T32 institutional training and F31 individual training grants.

**3.0 Results**

*3.1 Current Enrollments Across Programs*

The survey of data started with the required self-reporting data from all documented CAMPEP educational programs to understand the distribution of students across institutions and countries. This consisted of 56 programs were found, and at least 24 did not have complete data online, but the program chairs were contacted to be given a chance to provide this information.  The majority of those contacted provided the data required to complete this survey.  Of this total group, 46 programs had complete non-zero data, with our total survey consisting of 31 domestic USA programs, 13 Canadian programs and 2 Irish programs.  The numbers of students per year fluctuated significantly from year to year, so a mean of the available 7 years (2017-2023) is reported here, plotted in Figure 1(a) for US universities and Figure 1(b) for international universities.

*3.2 Annual Trends in Degrees Cumulative Across Medical Physics*

The temporal trends of MS, PhD, Certificate and DMP programs are from cumulative self-reported statistics, plotted for the available 7 years of data from 2017 to 2023. Data from 2023 was available for nearly all programs and updates from some programs were provided by email contact with the program directors.

For domestic US programs, the trend for MS graduate growth is apparent, with an average growth of 17.7 new MS graduates per year over this time, while there was a growth of PhD graduates was 3.6 per year.  Certificate programs added an average of 1 graduate per year while the DMP programs were unchanged in average numbers.  It might be also noted that the 2023 numbers could be an underestimation as several programs are not fully updated at the time of this survey.

For international programs, the MS growth is dominated by programs in Ireland, at National University of Ireland Galway (accredited in 2015) and University College of Dublin (accredited in 2020), which have been adding students with accreditation.  Growth in Canada appears to be more about the time at which they received CAMPEP accreditation than a trend in program acceptance changes and given the smaller numbers it is harder to assess a trend.

*3.3 Faculty involvement in Medical Physics Degree Programs*

The numbers of participating faculty were compiled and graphed based upon a grouping of their home department, for the cumulative CAMPEP accredited graduate programs, as shown in Figure 3.  The first two graphs are separated for (a) domestic US and (b) international programs irrespective of degree makeup.  Then the data for US domestic programs are shown sub-divided out for (c) predominantly MS programs (as determined by >2 MS graduates per year with more MS than PhD graduates) and (d) predominantly PhD programs (as determined by >2 PhD graduates per year with more PhD than MS graduates).

*3.4 Student Support Mechanisms in Accredited Medical Physics Degree Programs*

Data was made available from CAMPEP for cumulative analysis on self-reported program funding of students. This was found from a survey of the graduate program directors with a series of questions.

The data is not quantitative but rather lists ways that students are funded in each program, but without % of each nor specificity as to exactly which programs and what number of students. As a result, it is hard to make global conclusions on this data. Nonetheless, a cumulative graph showing the fractional components of support are shown in Figure 4 across the programs

### *3.4 NIH Funding Trends to Related Departments*

The annual trends in grant funding are tabulated in Figure 5 for three representative departments that are listed in NIH RePORTER, classified as (i) RADIATION-DIAGNOSTIC ONCOLOGY dept, (ii) ENGINEERING, and with (iii) SURGERY is included as a comparator department. The % of grants funded scales linearly with dollars of funding, and so the data is shown as % of grants funded on the y-axis. The three relevant data pools for education were assessed as (a) all NIH funding, (b) National Cancer Institute (NCI) funding, and (c) student training grants funding.

Figure 5(a) shows all NIH grants, with the obvious trend of large growth in Engineering from 0.5% of all NIH grants in 1995 to now 3% of all grants in 2024. Interestingly this appears stabilized in the most recent range of 2022-2024. As a comparison, the Radiation-Diagnostic/Oncology listing grants had smaller growth, from 0.8% in 1995 up to now 1.6% in 2024. As a comparator, Surgery grant funding went from 0.95% in 1995 to 1.35% in 2024. Interestingly the trend of all 3 groups is upward in the last 10 years, although the slope of Engineering (+0.13 %/year) is 4X higher than those for Radiation (0.038%/year) and Surgery (0.032%/year). This is a clear indicator of the trend in higher education to the expansion of biomedical engineering research programs across the country during this time. Presumably, medical physics research programs are embedded within the Radiation-Diagnostic/Oncology dept funding listing given that most faculty (see Fig 3) are embedded within these two departments (Radiation Oncology and Radiology). Thus, the research funding increase for medical physics programs during this time period must have been modest.

However, given that Medical Physics is heavily dominated by radiation systems used in cancer, the grants funded by NCI were also examined, as shown for the same three programs in Figure 4(b). In this case the funding for Radiation-Diagnostic/Oncology is considerably higher, at a stable average of about 3.8% of all NCI, but with a notable rise to 5.6% in 2024. In this case Surgery NCI funding was stable near 2% average, but again with a slight rise in the latter couple of years to 2.8% by 2024. Engineering NCI funding % was a slow continuous rise throughout this period, reaching 2.7% by 2024.

In Figure 4(c) the training grants are listed with institutionally obtained training grants T32 listed and individual-support F31 training grants. The funding growth in T32 programs is quite similar in the two groups of departments, Engineering and Radiation, with perhaps more variation in the latter. On average their trajectories do not look that different. However, in the F31 grants there has been a strong increase in Engineering students gaining this funding since about 2018, increasing from about 25 to about 50 today. The funding in the Radiation area has grown from near zero in 1995 to about 18 per year in 2024.

### **4.0 Discussion**

With the ongoing expansion of Medical Physics as a discipline, and the nature of the education programs feeding this being a niche area, it is critical to understand how education and training is affecting the growth of the field [2-4]. The institutional decision making by faculty largely determines the free market forces of education supply. The analysis of this public data provides a window into these trends.

*4.1 Graduate Student Numbers and Growth*

Based upon the data in Figure 1, there are a wide range of program sizes and MS/PhD ratios across institutions. About 17 institutions appear focused predominantly on the MS degree (as measured by >2 per year with more MS than PhD graduates), while 8 appear more focused on PhD program (as measured by >2 per year with more PhD than MS graduates). About 4 have an equal balance of MS and PhD graduates (Buffalo, Louisiana, Virginia & Wayne State). Distributions of MS and PhD across Canada and Ireland are harder to judge because they are more about the accreditation process and which institutions choose to join CAMPEP. But there are notably high numbers of MS students in a few institutions that have accredited with CAMPEP in this program area.

Perhaps the most important feature seen in this data is the temporal trend of MS graduates in the past 7 years. The numbers of MS graduates outpaced the number of PHD graduates in 2018 and today there are twice as many MS graduates as PhD graduates. This trend seems is continuing with a growth of about 17.7 new MS graduates each year as compared to a growth of only 3.6 additional PHDs per year, in the USA programs. The number of certificate graduates has grown to about 1 additional person per year while the DMP program has no statistical growth. So, certificate programs remain a small but growing segment of the CAMPEP education pathway, while DMP programs have stayed small but stable.

Internationally, similar but smaller trends are seen in the Canadian system, with MS growth occurring substantially in the last two years, now surpassing PhD programs. In Ireland, only MS programs are charted, and so this is likely more a limitation of the system and data available than anything else, but there is growth, doubling in number from 2017 to 2022.

Certificate programs are a small but growing part of the training landscape, as predicted by several people [12]. Their numbers are nearly insignificant as compared to MS and PhD graduate numbers but again, likely serve to diversify the backgrounds of trained clinical physicists. In comparison DMP program graduates are quite few in number and do not appear to be growing. This may be because of limitations in pathways to fund this type of program, and competition by MS programs which lead to a shorter timeline to residency with perhaps similar career trajectories [13].

This survey was not able to analyze the diversity in education programs as this type of demographic data is not publicly available from websites, however future studies might consider this as a needed value to ensure appropriate growth in the field [14].

*4.2 Departments Involved in Medical Physics Education*

The typical clinical departments involved are perhaps not surprising, with 30% Oncology, 28% Radiology, 8% Medicine & Surgery, 14% Medical Physics, and 11% Physics faculty, as shown in Figure 2(a). Given the employment of the vast number of clinical medical physicists are in Radiation Oncology followed by Radiology, these numbers seem reasonable. Participation of other departments can be for a variety of reasons to do with education focus, research capacity or institutional priorities in hiring faculty. In Ireland and Canada, they have a much smaller Radiology contingent, presumably illustrating more of a focus on therapeutic physics training.

When the USA numbers are broken into MS dominated (Fig 2c) versus PhD dominated (Fig 2d) programs, the MS dominated have a larger fraction of faculty in Oncology (40%) and Physics (26%), departments, while the PhD program dominated programs have a larger fraction of faculty in Radiology (40%) and Medical Physics (26%) departments.  There is also an interestingly large fraction from Medicine & Surgery in the PHD program group (12%).  These data may point to the fact that funding for PHD programs are more achieved through the Radiology, Med Phys or Medicine/Surgery faculty, while the Oncology & Physics faculty focus more on the revenue generating MS degrees for program support.  This is just speculation though.

*4.3 Student Funding Mechanisms*

Based upon the data in Figure 4, self-funding was the most commonly reported category from program directors in the USA, while in Canada and Ireland their portfolio of funding tools was more equally balanced across the list of options.  Self-funding by students in MS programs is the singular most dominant pathway for funding this education pathway, with these programs.  The motivations for using self-funding by students are likely mixed, but obviously are related to making the programs cost effective for the institution while serving the growing market need for more CAMPEP trained medical physicists.

Perhaps one of the most important observations in this data for the field of medical physics is the apparently very modest growth of both numbers of PhDs (Fig 1a) and of funding in Radiation Oncology and Radiology programs across the NIH (Fig 5a), at a time when growth in clinical medical physics is very strong.  There is some positive information in the NCI funding in Radiation-Diagnostic/Oncology has grown significantly in the last 5 years (Fig 5c), although the origins of this are not clear to date and may be in areas less related to medical physics, such as immunotherapy. The inclusion of engineering as a comparator program was intentional, because biomedical engineering and medical physics are so close to each other in terms of people skills and research topics.  The unparalleled growth in NIH funding within engineering schools is from a deep investment in research infrastructures and faculty across the country in the last several decades. Tenure track faculty who have  infrastructure, financial resources and a mandate and support to carry out research at 50% or greater FTE can obtain substantial NIH funding. This is the origin of NIH funding growth in engineering.  Importantly, this differentiation shows up in training grant data too, where the numbers are slowly rising and steady in Radiation-Diagnostic/Oncology but have a significant peak in recent years in engineering (Fig 5b).  More attention to training students and mentor faculty in grant writing to gain potential funding in F31s and T32s would benefit the field and provide needed funding support.

The dominant role of PhDs in all programs is to lead research and the stability of the number of PhDs is a good indicator that the foundations of this are still solid. However increasing numbers of MS students being trained and the significant increase in need for medical physicists nationwide likely means that research has been significantly decreasing as a fractional footprint of the field. This seems to match the trends in research versus clinical service across the country, as healthcare systems grow more in service than in research. Research has stayed more centralized to major academic medical centers that can afford to support the infrastructures and tenure track lines that support research careers and students.  It is also noteworthy that the funding in biomedical engineering has seen substantial growth, and it is likely that PhD growth is occurring much more outside of CAMPEP accredited programs.  There are T32 training programs in imaging sciences and biomedical imaging which are outside of the accredited prevue of CAMPEP.  Some analysis of this in future studies is warranted to see if the field of medical physics is losing its research footprint to non-medical physics programs.

CAMPEP survey data on funding mechanisms for student support was made available as well. This was survey data provided voluntarily from individual program directors, listing the general categories of funding mechanisms available to programs. While this data is really imperfect because it is not qualitative rather than quantitative, and there are overlapping and complimentary information sets, without definition of numbers of students nor percentages of the program, but still it was used as a tool to assay the funding of masters and doctoral programs as a pool.

**5.0 Summary**

This survey of public data provided a useful perspective on the status and trends in Medical Physics education. From 2017 to 2023 there has been consistent growth in MS graduates, increasing at an average of 17.7 per year, while the growth in PhDs has been slower at 3.6 per year. By 2024 there should be close to double the number of MS graduates as PhD graduates. As the field of medical physics grows, this survey suggests that growth in the field of medical physics education is and will be dominated by MS graduates. These programs which focus on larger MS graduating numbers have a large fraction of self-funded students in domestic programs. Interestingly, international MS and domestic PhD programs have a broader range of financial support mechanisms. Additionally, those programs dominated by MS education graduates have their largest fraction of faculty in Radiation Oncology departments followed by Physics departments, whereas those programs dominated by PhD graduates have their largest fraction of faculty in Radiology departments followed by Medical Physics departments. Overall NIH funding in the space of Radiation-Diagnostic/Therapy has been slowly growing over this timeframe, but with a notable recent rise in NCI funding in the last 5 years. However, there has been a substantial 5X-6X in NIH funding for engineering research programs during this time, including significant increases in F31 and T32 training grants. The origins of these trends are not entirely clear from this survey, but it seems as though the research funding that supports PhD training in medical physics is likely to be missing the ongoing NIH funding growth trend in biomedical engineering. The overall picture presented in this data is one where the field of medical physics is growing solidly as a clinical discipline, with a somewhat static research component in terms of size and support.

**Conflict of Interest**

The authors have no conflicts of interest relevant to the content of this paper.

**Figure Captions**

**Figure 1.** The average data over 5 years for MS and PhD graduating student numbers are shown here for (a) domestic USA institutions and (b) Canadian & Irish universities.

**Figure 2.** Annual trends shown for graduates from MS, PhD, Certificate and DMP programs for (a) domestic USA programs, (b) Canadian programs and (c) Irish programs.

**Figure 3.** The faculty involvement in programs were separated by their home departments, graphed as % pie charts for (a) domestic USA program faculty and (b) international Canadian & Irish program faculty. For the USA data, it was further separated into (c) those programs dominated by MS programs and (d) those dominated by PhD programs.

**Figure 4.** Listing of aggregate support mechanisms from CAMPEP program survey is shown, unweighted by any other factors, for (a) USA programs, (b) international programs.

**Figure 5.** Annual NIH funding trends reported by NIH RePORTER for three department categories, (i) Radiation-Diagnostic/Oncology, (ii) Engineering, and (iii) Surgery. These are graphed as (a) a % of all NIH grants, (b) a % of all NCI grants, and (c) for student training support grants, institutional T32 and individual F31.

**Figure 1**

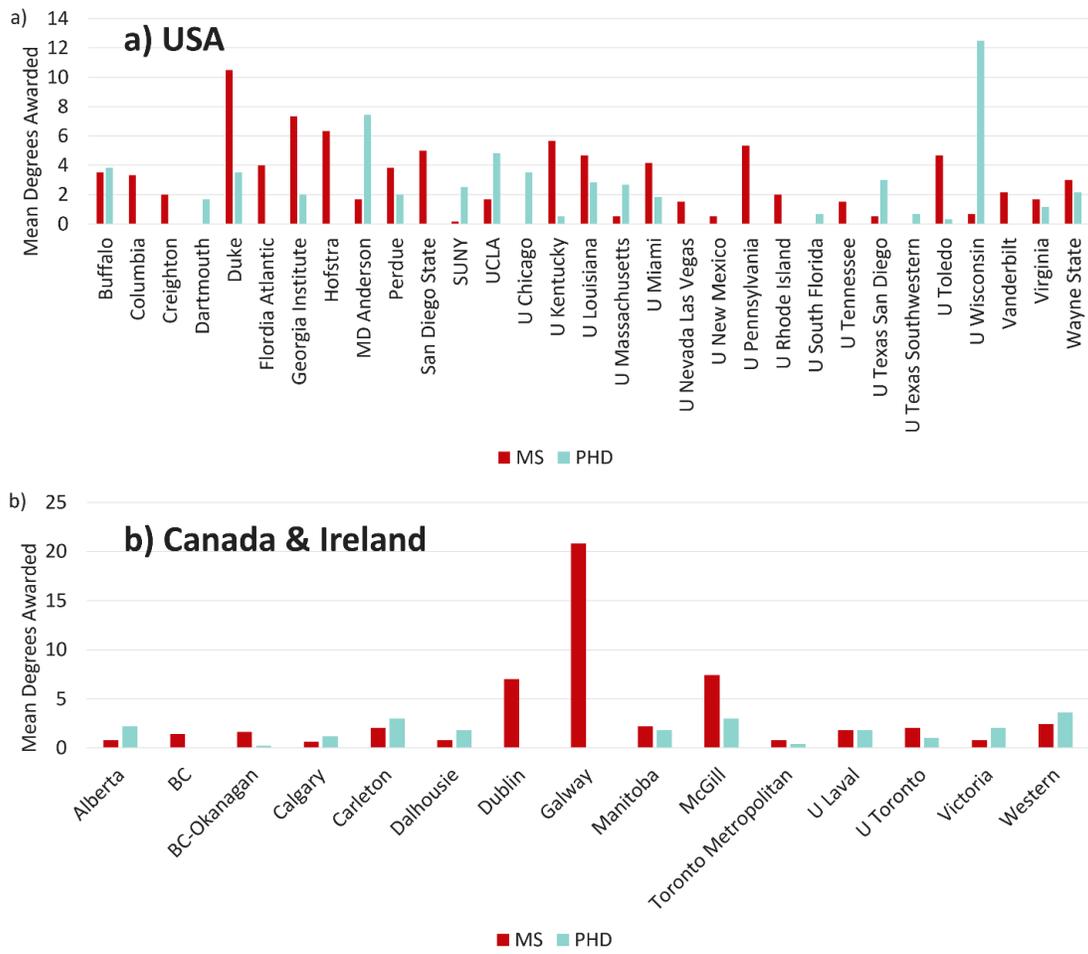

**Figure 2**

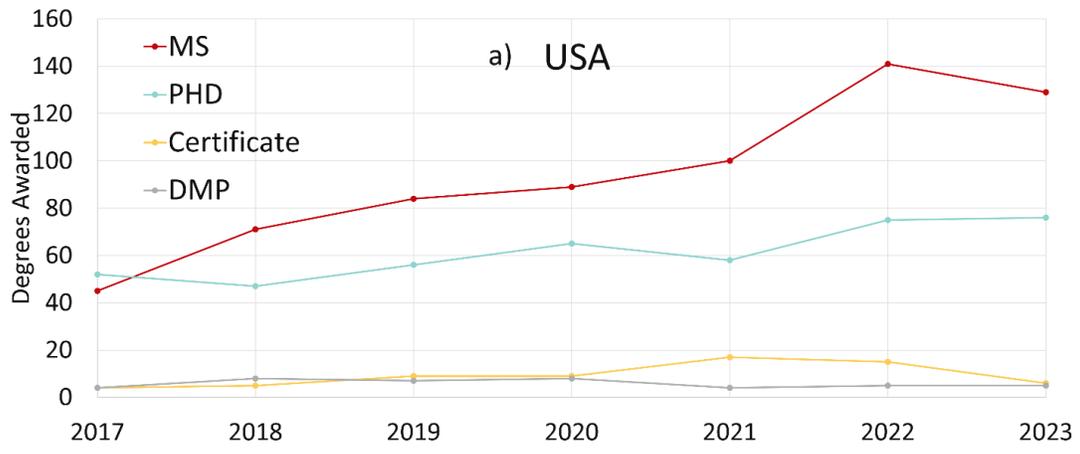
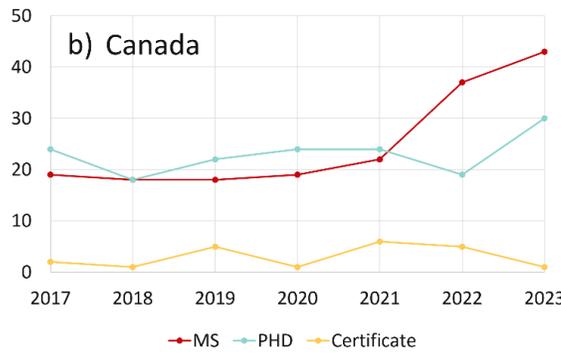
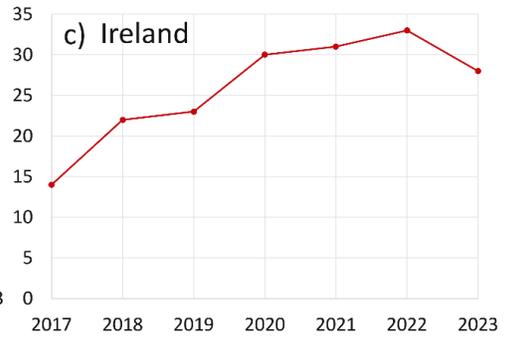

*Figure 3*

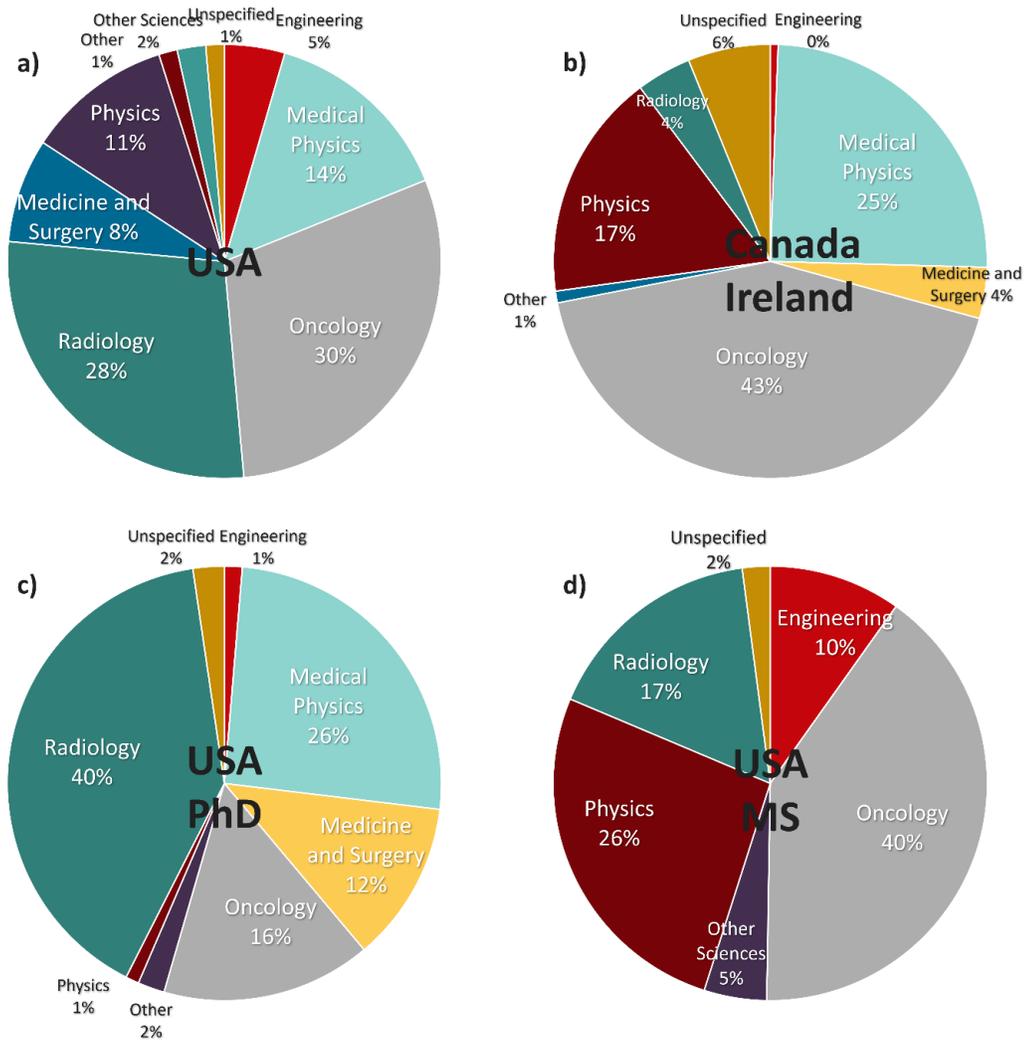

Figure 4

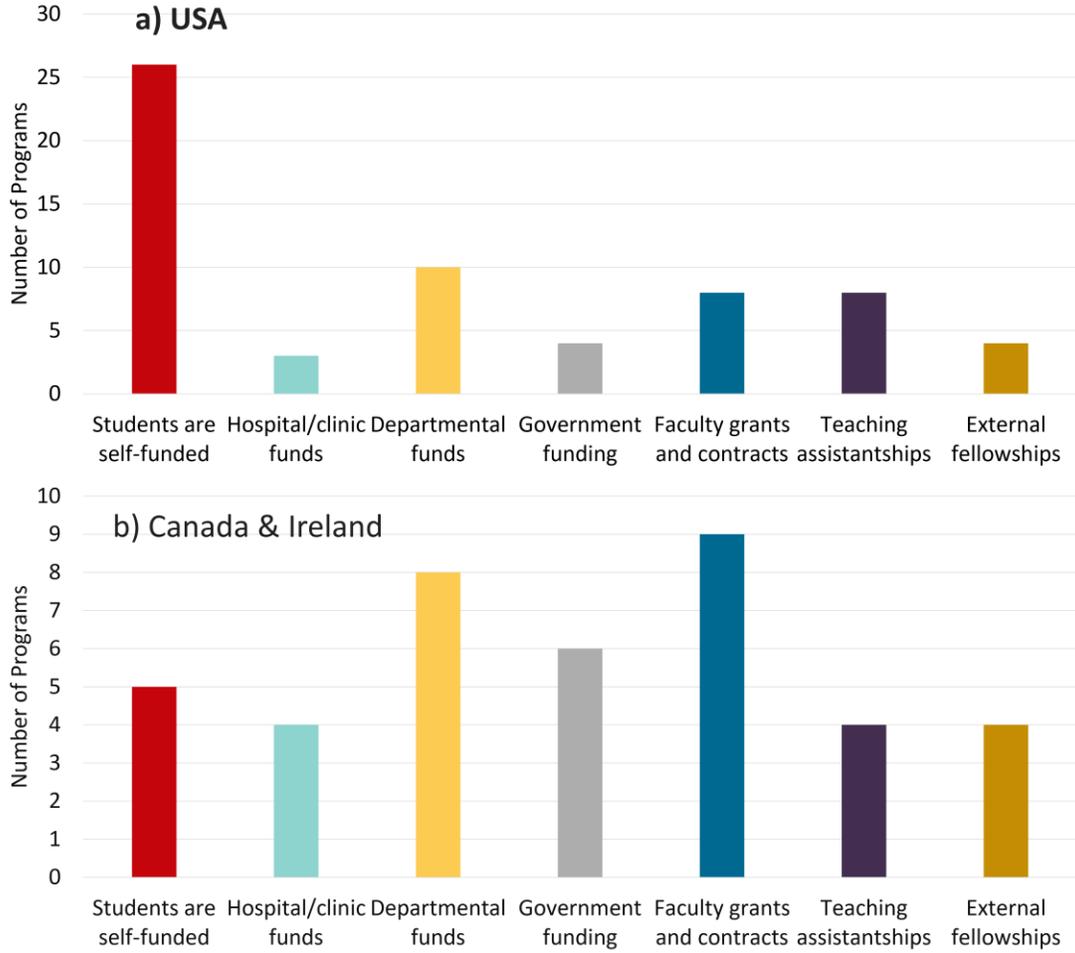

Figure 5

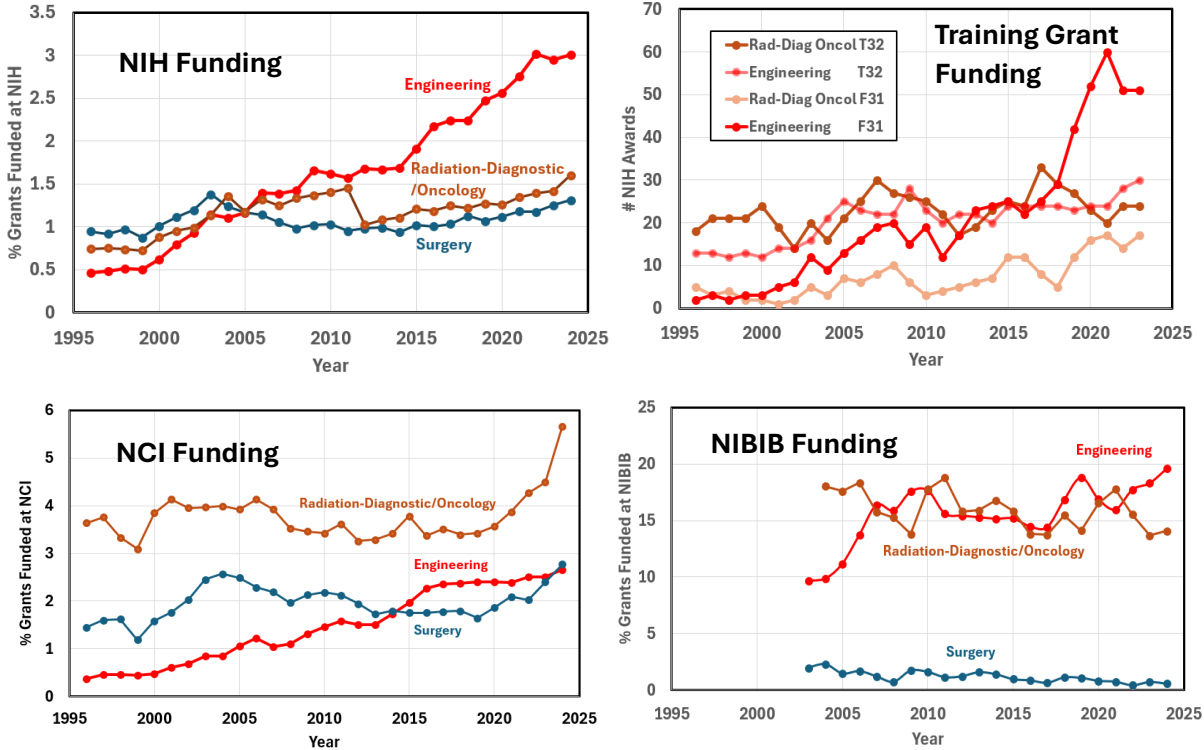